\documentclass[twocolumn,aps,prb]{revtex4}

\usepackage{graphicx}% Include figure files
\begin{document}

\title{Circuit Quantum Electrodynamics:\\
Coherent Coupling of a Single Photon to a Cooper Pair Box}

\author{A.~Wallraff}
\affiliation{Departments of Applied Physics and Physics, Yale
University, New Haven, CT 06520}
\author{D.~Schuster}
\affiliation{Departments of Applied Physics and Physics, Yale
University, New Haven, CT 06520}
\author{A.~Blais}
\affiliation{Departments of Applied Physics and Physics, Yale
University, New Haven, CT 06520}
\author{L.~Frunzio}
\affiliation{Departments of Applied Physics and Physics, Yale
University, New Haven, CT 06520}
\author{R.-S.~Huang}
\altaffiliation{Department of Physics, Indiana University,
Bloomington, IN 47405} \affiliation{Departments of Applied Physics
and Physics, Yale University, New Haven, CT 06520}
\author{J.~Majer}
\affiliation{Departments of Applied Physics and Physics, Yale
University, New Haven, CT 06520}
\author{S.~Kumar}
\affiliation{Departments of Applied Physics and Physics, Yale
University, New Haven, CT 06520}
\author{S.~M.~Girvin}
\affiliation{Departments of Applied Physics and Physics, Yale
University, New Haven, CT 06520}
\author{R.~J.~Schoelkopf}
\affiliation{Departments of Applied Physics and Physics, Yale
University, New Haven, CT 06520}
\date{\today, accepted for publication in \emph{Nature (London)}}

\def\be{\begin{equation}}
\def\ee{\end{equation}}
\def\omegar{\omega_{\rm r}}
\def\ncav{n_{\rm cav}}
\def\ndet{n_{\rm amp}}
\def\ncrit{n_{\rm crit}}

\begin{abstract}
Under appropriate conditions, superconducting electronic circuits
behave quantum mechanically, with properties that can be designed
and controlled at will. We have realized an experiment in which a
superconducting two-level system, playing the role of an
artificial atom, is strongly coupled to a single photon stored in
an on-chip cavity. We show that the atom-photon coupling in this
circuit can be made strong enough for coherent effects to dominate
over dissipation, even in a solid state environment. This new
regime of matter light interaction in a circuit can be exploited
for quantum information processing and quantum communication. It
may also lead to new approaches for single photon generation and
detection.
\end{abstract}

\maketitle

The interaction of matter and light is one of the fundamental
processes occurring in nature. One of its most elementary forms is
realized when a single atom interacts with a single photon.
Reaching this regime has been a major focus of research in atomic
physics and quantum optics\cite{Walls94} for more than a decade
and has created the field of cavity quantum
electrodynamics\cite{Mabuchi02} (CQED). In this article we show
for the first time that this regime can be realized in a solid
state system, where we have experimentally observed the coherent
interaction of a superconducting two-level system with a single
microwave photon. This new paradigm of \emph{circuit quantum
electrodynamics} may open many new possibilities for studying the
strong interaction of light and matter.

In atomic cavity QED an isolated atom with electric dipole moment
$d$ interacts with the vacuum state electric field $E_0$ of a
cavity. The quantum nature of the electromagnetic field gives rise
to coherent oscillations of a single excitation between the atom
and the cavity at the vacuum Rabi frequency $\nu_{\rm{Rabi}} = 2 d
E_0/h$, which can be observed experimentally when
$\nu_{\rm{Rabi}}$ exceeds the rates of relaxation and decoherence
of both the atom and the field. This effect has been observed both
in the time domain using large dipole moment Rydberg atoms in 3D
microwave cavities \cite{raimond01} and spectroscopically using
alkali atoms in very small optical cavities with large vacuum
fields \cite{Thompson92,Hood02}.

Our experimental implementation of cavity quantum electrodynamics
in a circuit consists of a fully electrically controllable
superconducting quantum two-level system, the Cooper pair
box\cite{Bouchiat98}, coupled to a single mode of the quantized
radiation field in an on-chip cavity formed by a superconducting
transmission line resonator. Coherent quantum effects have been
recently observed in several superconducting circuits
\cite{Nakamura99,Vion02,Martinis02,Yu02,Chiorescu03,Yamamoto03},
making these systems well suited for use as quantum bits (qubits)
for quantum information processing \cite{Bennett00,Nielsen00}. Of
these superconducting qubits, the Cooper pair box is especially
well suited for cavity QED because of its large effective electric
dipole moment $d$, which can be made more than $10^4$ times larger
than in an alkali atom in the ground state and still $10$ times
larger than in a typical Rydberg atom. In addition, the small size
of the quasi one-dimensional transmission line cavity used in our
experiments generates a large vacuum electric field $E_0$. As
suggested in our earlier theoretical study \cite{Blais04}, the
simultaneous combination of large dipole moment and large field
strength in our implementation is ideal for reaching the strong
coupling limit of cavity QED in a circuit. Other solid state
analogs of strong coupling cavity QED have been envisaged by many
authors in superconducting
\cite{Makhlin01,Buisson01,Marquardt01,Al-Saidi01,Plastina03,Blais03,Yang03,you03a},
semiconducting \cite{Kiraz01,Childress04} and even
micro-mechanical systems \cite{Armour02,Irish03}. First steps
towards realizing such a regime have been made for semiconductors
\cite{Weissbuch92,Kiraz01,Vuckovic03}. Our experiments constitute
the first experimental observation of strong coupling cavity QED
with a \emph{single} artificial atom and a \emph{single} photon in
a solid state system.

\begin{figure*}[tbp]
\begin{center}
\includegraphics[width = 1.5\columnwidth]{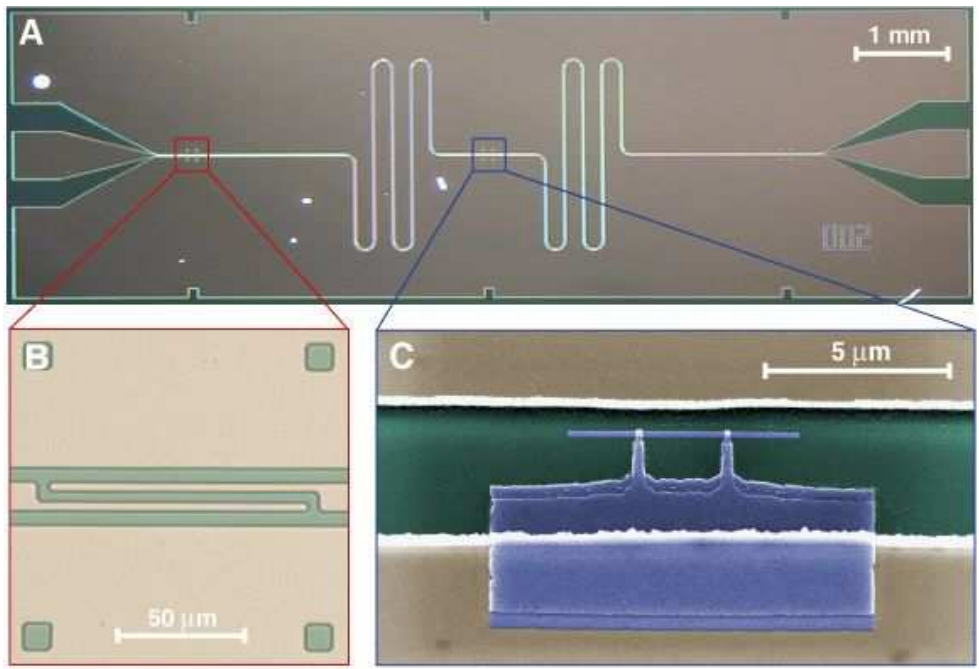}
\end{center}
\caption{Integrated circuit for cavity QED. \textbf{a} The
superconducting niobium coplanar waveguide resonator is fabricated
on an oxidized $10\times3 \, \rm{mm}^2$ silicon chip using optical
lithography. The width of the center conductor is $w = 10 \,
\rm{\mu m}$ separated from the lateral ground planes extending to
the edges of the chip by a gap of width $d = 5 \, \rm{\mu m}$
resulting in a wave impedance of the structure of $Z = 50 \,
\rm{\Omega}$ being optimally matched to conventional microwave
components. The length of the meandering resonator is $l = 24 \,
\rm{mm}$. It is coupled by a capacitor at each end of the
resonator (see b) to an input and output feed line, fanning out to
the edge of the chip and keeping the impedance constant.
\textbf{b} The capacitive coupling to the input and output lines
and hence the coupled quality factor $Q$ is controlled by
adjusting the length and separation of the finger capacitors
formed in the center conductor. \textbf{c} False color electron
micrograph of a Cooper pair box (blue) fabricated onto the silicon
substrate (green) into the gap between the center conductor (top)
and the ground plane (bottom) of a resonator (beige) using
electron beam lithography and double angle evaporation of
aluminum. The Josephson tunnel junctions are formed at the overlap
between the long thin island parallel to the center conductor and
the fingers extending from the much larger reservoir coupled to
the ground plane.} \label{fig:sample}
\end{figure*}

\section{The cavity QED circuit}
The on-chip cavity is realized as a distributed transmission line
resonator patterned into a thin superconducting film deposited on
the surface of a silicon chip. The quasi-one-dimensional coplanar
waveguide resonator \cite{Day03} consists of a narrow center
conductor of length $l$ and two nearby lateral ground planes, see
Fig.~\ref{fig:sample}a. Its full wave resonance occurs at a
frequency of $\nu_{\rm{r}}= 6.04 \, \rm{GHz}$. Close to its
resonance frequency $\omega_{\rm{r}} = 2 \pi \nu_{\rm{r}} =
1/\sqrt{LC}$ the resonator can be modelled as a parallel
combination of a capacitor $C$ and an inductor $L$ (a resistor $R$
can be omitted since the internal losses are small). This simple
resonant circuit behaves as a harmonic oscillator described by the
Hamiltonian $H_{\rm{r}} = \hbar \omega_{\rm{r}} (a^{\dagger}a +
1/2)$, where $a^{\dagger}$ ($a$) is the creation (annihilation)
operator for a single photon in the electromagnetic field and
$\langle a^{\dagger}a \rangle = \langle \widehat{n} \rangle = n$
is the average photon number. Cooling such a resonator down to
temperatures well below $T^{\star} = \hbar \omega_{\rm{r}}/k_B
\approx 300 \, \rm{mK}$, the average photon number $n$ can be made
very small. At our operating temperatures of $T < 100 \, \rm{mK}$
the thermal occupancy is $n < 0.06$, so the resonator is near its
quantum ground state. The vacuum fluctuation energy stored in the
resonator in the ground state gives rise to an effective rms
voltage of $V_{\rm{rms}} = \sqrt{\hbar \omega_{\rm{r}}/{2C}}
\approx 1 \, \rm{\mu V}$ on its center conductor. The magnitude of
the electric field between the center conductor and the ground
plane in the vacuum state is then a remarkable $E_{\rm{rms}}
\approx 0.2 \, \rm{V/m}$, which is some $100$ times larger than in
atomic microwave cavity QED\cite{raimond01}. The large vacuum
field strength resulting from the extremely small effective mode
volume ($\sim 10^{-6}$ cubic wavelengths) of the resonator is
clearly advantageous for cavity QED not only in solid state but
also in atomic systems \cite{Sorensen04}. In our experiments this
vacuum field is coupled capacitively to a superconducting qubit.
At the full wave resonance the electric field has antinodes at
both ends and at the center of the resonator and its magnitude is
maximal in the gap between the center conductor and the ground
plane where we place our qubit.

\begin{figure*}[tbp]
\begin{center}
\includegraphics[width=2.0\columnwidth]{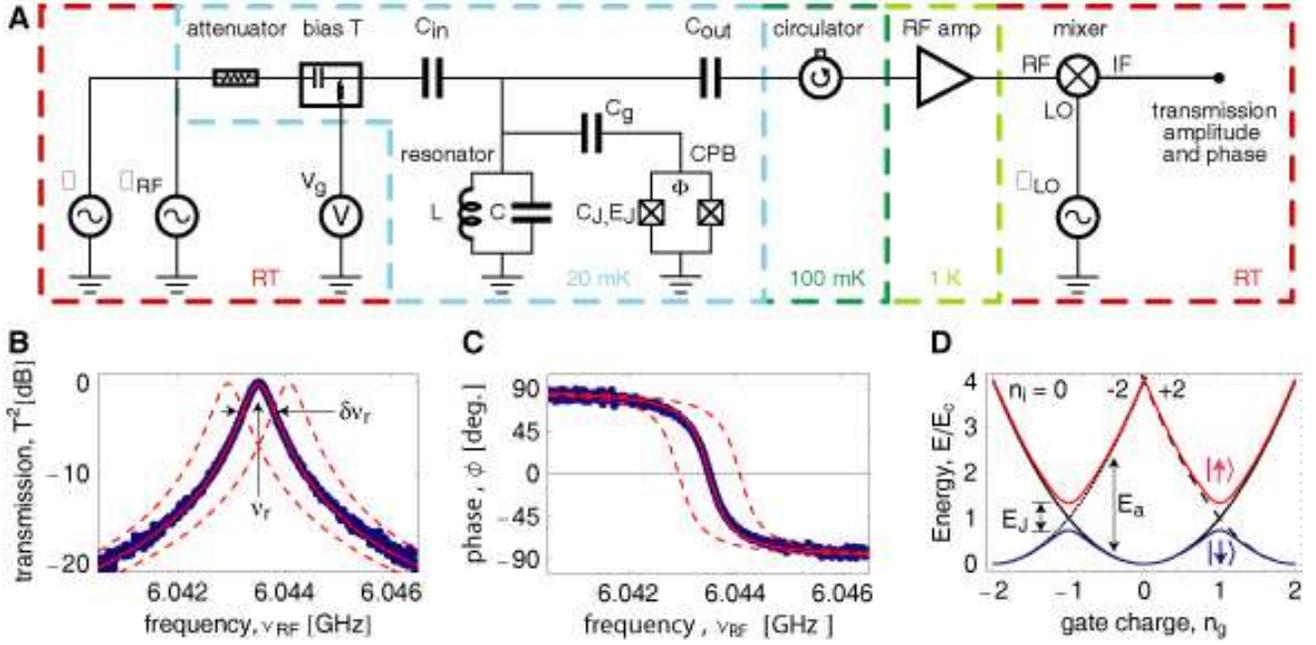}
\end{center}
\caption{Measurement scheme, resonator and Cooper pair box.
\textbf{a} The resonator with effective inductance $L$ and
capacitance $C$ coupled through the capacitor $C_g$ to the Cooper
pair box with junction capacitance $C_J$ and Josephson energy
$E_J$ forms the circuit QED system which is coupled through
$C_{\rm{in/out}}$ to the input/ouput ports. The value of $E_J$ is
controllable by the magnetic flux $\Phi$. The input microwaves at
frequencies $\omega$ and $\omega_{\rm{RF}}$ are added to the gate
voltage $V_g$ using a bias-tee. After the transmitted signal at
$\omega_{\rm{RF}}$ is amplified using a cryogenic high electron
mobility (HEMT) amplifier and
% with a noise temperature of $T_{\rm{N}}\sim 6 \, \rm{K}$
mixed with the local oscillator at $\omega_{\rm{LO}}$, its
amplitude and phase are determined. The circulator and the
attenuator prevent leakage of thermal radiation into the
resonator. The temperature of individual components is indicated.
\textbf{b} Measured transmission power spectrum of the resonator
(blue dots), the full line width $\delta \nu_r$ at half maximum
and the center frequency $\nu_r$ are indicated. The solid red line
is fit to a Lorentzian with $Q = \nu_r/\delta\nu_r\approx 10^4$.
\textbf{c} Measured transmission phase $\phi$ (blue dots) with fit
(red line). In panels b and c the dashed lines are theory curves
shifted by $\pm \delta\nu_r$ with respect to the data. \textbf{d}
Energy level diagram of a Cooper pair box. The electrostatic
energy $(n_i-n_g)^2e^2/2C_{\Sigma}$ is indicated for $n_i = 0$
(solid black line), $-2$ (dotted line) and $+2$ (dashed line)
excess electrons forming Cooper pairs on the island.
$C_{\rm{\Sigma}}$ is the total capacitance of the island given by
the sum of the capacitances $C_J$ of the two tunnel junctions, the
coupling capacitance $C_g$ to the center conductor of the
resonator and any stray capacitances. In such a system $n_i$ is
well determined if $E_C/k_B$ is much larger than the bath
temperature $T$ and if the quantum fluctuations induced by the
tunnelling of electrons through the junctions are small. In the
absence of Josephson tunnelling the states with $n_i$ and $n_i +
2$ electrons on the island are degenerate at $n_g = 1$. The
Josephson coupling mediated by the weak link formed by the tunnel
junctions between the superconducting island and the reservoir
lifts this degeneracy and opens up a gap proportional to the
Josephson energy $E_J$. A ground state band
$\left|\downarrow\right\rangle$ and an excited state band
$\left|\uparrow\right\rangle$ are formed with a gate charge and
flux bias dependent energy level separation of $E_a$.}
\label{fig:circuit}
\end{figure*}

The resonator is coupled via two coupling capacitors
$C_{\rm{in/out}}$, one at each end, see Fig.~\ref{fig:sample}b, to
the input and output transmission lines through which its
microwave transmission can be probed, see Fig.~\ref{fig:circuit}a.
The predominant source of dissipation is the loss of photons from
the resonator through these input and output ports at a rate
$\kappa = \omega_r/Q$, where $Q$ is the (loaded) quality factor of
the resonator. The internal (uncoupled) loss of the resonator is
negligible ($Q_{\rm{int}} \approx 10^6$). Thus, the average photon
lifetime in the resonator $T_r = 1/\kappa$ exceeds $100 \,
\rm{ns}$ even for moderate quality factors of $Q \approx 10^4$.

Our choice of artificial atom is a mesoscopic superconducting
electrical circuit, the Cooper pair box\cite{Bouchiat98} (CPB),
the quantum properties of which have been clearly demonstrated
\cite{Nakamura99,Makhlin01,Lehnert03}. The Cooper pair box
consists of a several micron long and sub-micron wide
superconducting island which is coupled via two sub-micron size
Josephson tunnel junctions to a much larger superconducting
reservoir. We have fabricated such a structure into the gap
between the center conductor and the ground plane in the center
section of the coplanar waveguide resonator, see
Fig.~\ref{fig:sample}c.

The Cooper Pair box is a tunable two-state system described by the
Hamiltonian $H_a = - 1/2 \left(E_{\rm{el}} \, \sigma_x + E_J \,
\sigma_z\right)$, where $E_{\rm{el}}$ is the electrostatic energy
and $E_J$ is the Josephson energy of the circuit. The
electrostatic energy $E_{\rm{el}} = 4 E_C \left(1-n_g\right)$ is
proportional to the charging energy $E_{\rm{C}} =
e^2/2C_{\rm{\Sigma}}$, where $C_{\rm{\Sigma}}$ is the total
capacitance of the island. $E_{\rm{el}}$ can be tuned by a gate
charge $n_g = V_g C^{\star}_g/e$, where $V_g$ is the voltage
applied to the input port of the resonator and $C^{\star}_g$ the
effective dc-capacitance between the input port of the resonator
and the island of the Cooper pair box. The Josephson energy $E_J =
E_{J,\rm{max}} \cos\left(\pi\Phi_b\right)$ is proportional to
$E_{\rm{J,max}} = h \Delta/8e^2R_J$, where $\Delta$ is the
superconducting gap and $R_J$ is the tunnel junction resistance.
Using an external coil, $E_J$ can be tuned by applying a flux bias
$\Phi_b = \Phi/\Phi_0$ to the loop formed by the two tunnel
junctions, the island and the reservoir. We denote the ground
state of this system as $\left|\downarrow\right\rangle$ and the
excited state as $\left|\uparrow\right\rangle$, see
Fig.~\ref{fig:circuit}d. The energy level separation between these
two states is then given by $E_a = \hbar \omega_a =
\sqrt{E_{J,\rm{max}}^2\cos^2(\pi\Phi_b)+16
E_C^2\left(1-n_g\right)^2}$.

The Hamiltonian $H_a$ of the Cooper pair box can be readily
engineered by choosing the parameters $E_{J,\rm{max}}$ and $E_C$
during fabrication and by tuning the gate charge $n_g$ and the
flux bias $\Phi_b$ in situ during experiment. The quantum state of
the Cooper pair box can be manipulated using microwave pulses
\cite{Nakamura99,Nakamura02}. Coherence in the Cooper pair box is
limited both by relaxation from the excited state
$\left|\uparrow\right\rangle$ into the ground state
$\left|\downarrow\right\rangle$ at a rate $\gamma_1$ corresponding
to a lifetime of $T_1 = 1/\gamma_1$ and by fluctuations in the
level separation giving rise to dephasing at a rate
$\gamma_{\varphi}$ corresponding to a dephasing time
$T_{\varphi}=1/\gamma_{\varphi}$. The total decoherence rate
$\gamma = \gamma_1/2 + \gamma_{\varphi}$ has contributions from
the coupling to fluctuations in the electromagnetic environment
\cite{Lehnert03,Schoelkopf02} but is possibly dominated by
non-radiative processes. Relevant sources of dephasing include
coupling to low frequency fluctuations of charges in the substrate
or tunnel junction barrier and coupling to flux motion in the
superconducting films.

\section{The qubit-cavity coupling}

The coupling between the radiation field stored in the resonator
and the Cooper pair box is realized through the coupling
capacitance $C_g$. A voltage $V_{\rm{rms}}$ on the center
conductor of the resonator changes the energy of an electron on
the island of the Cooper pair box by an amount $\hbar g = e
V_{\rm{rms}} C_g/C_{\Sigma}$, where $g$ is the coupling strength.
The energy $\hbar g$ is the dipole interaction energy $d E_0$
between the qubit and the vacuum field in the resonator. We have
shown\cite{Blais04} that the Hamiltonian describing this coupled
system has the form $H_{\rm{JC}} = H_r + H_a + \hbar g
(a^{\dagger}\sigma^- + a\sigma^{+})$, where $\sigma^+$
($\sigma^-$) creates (annihilates) an excitation in the Cooper
pair box. The Jaynes-Cummings Hamiltonian, $H_{\rm{JC}}$,
describing our coupled circuit is well known from cavity QED. It
describes the coherent exchange of energy between a quantized
electromagnetic field and a quantum two-level system at a rate
$g/2\pi$, which is observable if $g$ is much larger than the
decoherence rates $\gamma$ and $\kappa$. This situation is
achieved in our experiments and is referred to as the strong
coupling limit ($g>\left[\gamma,\kappa\right]$) of cavity
QED\cite{Haroche92}. If such a system is prepared with the
electromagnetic field in its ground state $\left|0\right\rangle$,
the two-level system in its excited state
$\left|\uparrow\right\rangle$ and the two constituents are in
resonance $\omega_a = \omega_r$, it oscillates between states
$\left|0,\uparrow\right\rangle$ and
$\left|1,\downarrow\right\rangle$ at the vacuum Rabi frequency
$\nu_{\rm{Rabi}}=g/\pi$. In this situation the eigenstates of the
coupled system are symmetric and antisymmetric superpositions of a
single photon in a resonator and an excitation in the Cooper pair
box $\left|\pm\right\rangle = \left(\left|0,\uparrow\right\rangle
\pm \left|1,\downarrow\right\rangle\right)/\sqrt{2}$ with energies
$E_{\pm} = \hbar \left(\omega_{r} \pm g\right)$. Although the
eigenstates $\left|\pm\right\rangle$ are entangled, their
entangled character is not addressed in our current CQED
experiment. In our experiment, it is the energies $E_{\pm}$ of the
coherently coupled system which are probed spectroscopically.

In the non-resonant (`dispersive') case the qubit is detuned by an
amount $\Delta = \omega_a - \omega_r$ from the cavity resonance.
The strong coupling between the field in the resonator and the
Cooper pair box can be used to perform a quantum nondemolition
(QND) measurement of the quantum state of the Cooper pair box. For
$\left | \Delta \right | \gg g$, diagonalization of the coupled
quantum system leads to the effective Hamiltonian\cite{Blais04}
$$H \approx \hbar \left(\omega_r +
\frac{g^2}{\Delta}\sigma_z\right)a^{\dagger}a + \frac{1}{2}\hbar
\left(\omega_a+\frac{g^2}{\Delta}\right)\sigma_z \, ,$$ where the
terms proportional to $a^{\dagger}a$ determine the resonator
properties and the terms proportional to $\sigma_z$ determine the
circuit properties. It is easy to see that the transition
frequency of the resonator $\omega_r \pm g^2/\Delta$ is
conditioned by the qubit state $\sigma_z = \pm 1$ and depends on
the coupling strength $g$ and the detuning $\Delta$. Thus by
measuring the transition frequency of the resonator the qubit
state can be determined. Similarly, the level separation in the
qubit $\Delta E_a = \hbar \left(\omega_a + 2 a^{\dagger}a \,
g^2/\Delta + g^2/\Delta\right)$ is conditioned by the number of
photons in the resonator. The term $2 \, a^{\dagger}a \,
g^2/\Delta$ linear in the photon number $\hat{n}$ is known as the
ac-Stark shift and the term $g^2/\Delta$ as the Lamb shift as in
atomic physics. All terms in this Hamiltonian, with the exception
of the Lamb shift, are clearly identified in the results of our
circuit QED experiments.

\section{The measurement technique}
The properties of this strongly coupled circuit QED system are
determined by probing the level separation of the resonator states
spectroscopically\cite{Blais04}. The amplitude $T$ and phase
$\phi$ of a microwave probe beam of power $P_{\rm{RF}}$
transmitted through the resonator are measured versus probe
frequency $\omega_{\rm{RF}}$. A simplified schematic of the basic
microwave circuit is shown in Fig.~\ref{fig:circuit}a. In this
setup the Cooper pair box acts as a circuit element that has an
effective capacitance which is dependent on its $\sigma_z$
eigenstate, the coupling strength $g$, and its detuning $\Delta$.
It is connected through $C_g$ in parallel with the resonator. This
variable capacitance changes the resonance frequency of the
resonator which can be probed by measuring its microwave
transmission spectrum. The transmission $T^2$ and phase $\phi$ of
the resonator for a far detuned qubit ($g^2/\Delta\kappa \ll 1$),
i.e.~when the qubit is effectively decoupled from the resonator,
are shown in Figs.~\ref{fig:circuit}b and c. In this case, the
transmission is a Lorentzian line of width $\delta\nu_r = \nu_r/Q$
at the resonance frequency $\nu_r$ of the bare resonator. The
transmission phase $\phi$ displays a step of $\pi$ at $\nu_r$ with
a width determined by $Q$. The expected transmission spectrum in
both phase and amplitude for a frequency shift $\pm g^2/\Delta$ of
one resonator line width are shown by dashed lines in
Fig.~\ref{fig:circuit}b and c. Such small shifts in the resonator
frequency are sensitively measured in our experiments using probe
beam powers $P_{\rm{RF}}$ which controllably populate the
resonator with average photon numbers from $n \approx 10^3$ down
to the sub photon level $n \ll 1$. It is worth noting that in this
architecture both the resonator and the qubit can be controlled
and measured using capacitive and inductive coupling only,
i.e.~without attaching any dc-connections to either system.

\begin{figure*}[tbp]
\begin{center}
\includegraphics[width = 1.7\columnwidth]{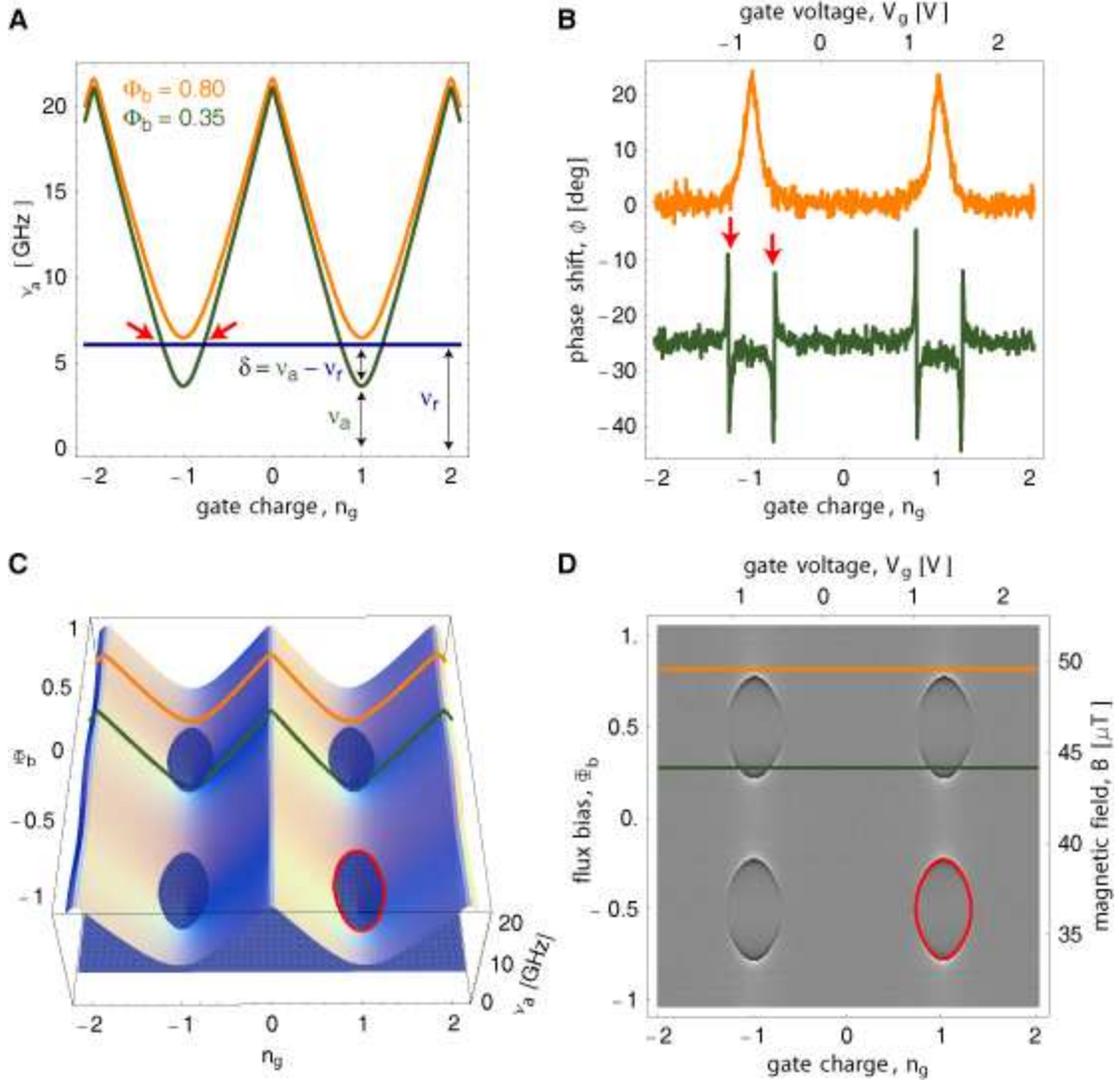}
\end{center}
\caption{Strong coupling circuit QED in the dispersive regime.
\textbf{a} Calculated level separation $\nu_a = \omega_a/2\pi =
E_a/h$ between ground $\left|\downarrow\right\rangle$ and excited
state $\left|\uparrow\right\rangle$ of qubit for two values of
flux bias $\Phi_b = 0.8$ (orange line) and $\Phi_b = 0.35$ (green
line). The resonator frequency $\nu_r = \omega_r/2\pi$ is shown by
a blue line. Resonance occurs at $\nu_a = \nu_r$ symmetrically
around degeneracy $n_g=1$, see red arrows. The detuning
$\Delta/2\pi = \delta = \nu_a-\nu_r$ is indicated. \textbf{b}
Measured phase shift $\phi$ of the transmitted microwave for
values of $\Phi_b$ in a. Green curve is offset by $-25 \,
\rm{deg}$ for visibility. \textbf{c} Calculated qubit level
separation $\nu_a$ versus bias parameters $n_g$ and $\Phi_b$. The
resonator frequency $\nu_r$ is indicated by the blue plane. At the
intersection, also indicated by the red curve in the lower right
hand quadrant, resonance between the qubit and the resonator
occurs ($\delta = 0$). For qubit states below the resonator plane
the detuning is $\delta<0$, above $\delta>0$. \textbf{d} Density
plot of measured phase shift $\phi$ versus $n_g$ and $\Phi_b$.
Light colors indicate positive $\phi$ ($\delta>0$), dark colors
negative $\phi$ ($\delta<0$). The red line is a fit of the data to
the resonance condition $\nu_a=\nu_r$. In c and d, the line cuts
presented in a and b are indicated by the orange and the green
line, respectively. The microwave probe power $P_{\rm{RF}}$ used
to acquire the data is adjusted such that the maximum intra
resonator photon number $n$ at $\nu_r$ is about 10 for
$g^2/\Delta\kappa \ll 1$. The calibration of the photon number has
been performed in situ by measuring the ac-Stark shift of the
qubit levels \cite{schuster04}.} \label{fig:footballs}
\end{figure*}

\section{The dispersive regime}
In the dispersive regime, the qubit is detuned from the resonator
by an amount $\Delta$ and thus induces a shift $g^2/\Delta$ in the
resonance frequency which we measure as a phase shift
$\tan^{-1}\left(2g^2/\kappa\Delta\right)$ of the transmitted
microwave at a fixed probe frequency $\omega_{\rm{RF}}$. The
detuning $\Delta$ is controlled \textsl{in situ} by adjusting the
gate charge $n_g$ and the flux bias $\Phi_b$ of the qubit. For a
Cooper pair box with Josephson energy $E_{J,\rm{max}}/h > \nu_r$
two different cases can be identified. In the first case, for bias
fluxes such that $E_J(\Phi_b)/h>\nu_r$, the qubit does not come
into resonance with the resonator for any value of gate charge
$n_g$, see Fig.~\ref{fig:footballs}a. As a result, the measured
phase shift $\phi$ is maximum for the smallest detuning $\Delta$
at $n_g=1$ and gets smaller as the detuning $\Delta$ increases,
see Fig.~\ref{fig:footballs}b. Moreover, $\phi$ is periodic in
$n_g$ with a period of $2e$, as expected. In the second case, for
values of $\Phi_b$ resulting in $E_J/h<\nu_r$, the qubit goes
through resonance with the resonator at two values of $n_g$. Thus,
the phase shift $\phi$ is largest as the qubit approaches
resonance ($\Delta \rightarrow 0$) at the points indicated by red
arrows, see Fig.~\ref{fig:footballs}b. As the qubit goes through
resonance, the phase shift $\phi$ changes sign as $\Delta$ changes
sign. This behavior is in perfect agreement with our predictions
based on the analysis of the circuit QED Hamiltonian in the
dispersive regime.

In Fig.~\ref{fig:footballs}c the qubit level separation $\nu_a =
E_a/h$ is plotted versus the bias parameters $n_g$ and $\Phi_b$.
The qubit is in resonance with the resonator ($\nu_a = \nu_r$) at
the set of parameters $[n_g,\Phi_b]$ also indicated by the red
curve in one quadrant of the plot. The measured phase shift $\phi$
is plotted versus both $n_g$ and $\Phi_b$ in
Fig.~\ref{fig:footballs}d. We observe the expected periodicity in
flux bias $\Phi_b$ with one flux quantum $\Phi_0$. The set of
parameters $[n_g,\Phi_b]$ for which the resonance condition is met
is marked by a sudden sign change in $\phi$. From this set, the
Josephson energy $E_{J,\rm{max}}$ and the charging energy $E_C$ of
the qubit can be determined in a fit to the resonance condition.
Using the measured resonance frequency $\nu_r = 6.044 \,
\rm{GHz}$, we extract the qubit parameters $E_{J,\rm{max}} = 8.0
\, (\pm 0.1) \, \rm{GHz}$ and $E_C = 5.2 \, (\pm 0.1) \,
\rm{GHz}$.

This set of data clearly demonstrates that the properties of the
qubit in this strongly coupled system can be determined in a
transmission measurement of the resonator and that full in situ
control over the qubit parameters is achieved. We note that in the
dispersive regime this new read-out scheme for the Cooper pair box
is most sensitive at charge degeneracy ($n_g=1$), where the qubit
is to first order decoupled from $1/f$ fluctuations in its charge
environment, which minimizes dephasing.  This property is
advantageous for quantum control of the qubit at $n_g=1$, a point
where traditional electrometry, using an SET for
example\cite{Lehnert03}, is unable to distinguish the qubit
states. Moreover, we note that this dispersive measurement scheme
performs a quantum non-demolition read-out of the qubit state
\cite{Blais04} which is the complement of the atomic microwave
cQED measurement in which the state of the photon field is
inferred non-destructively from the phase shift in the state of
atoms sent through the cavity \cite{Nogues99,raimond01}.

\begin{figure*}[tbp]
\begin{center}
\includegraphics[width=2.0\columnwidth]{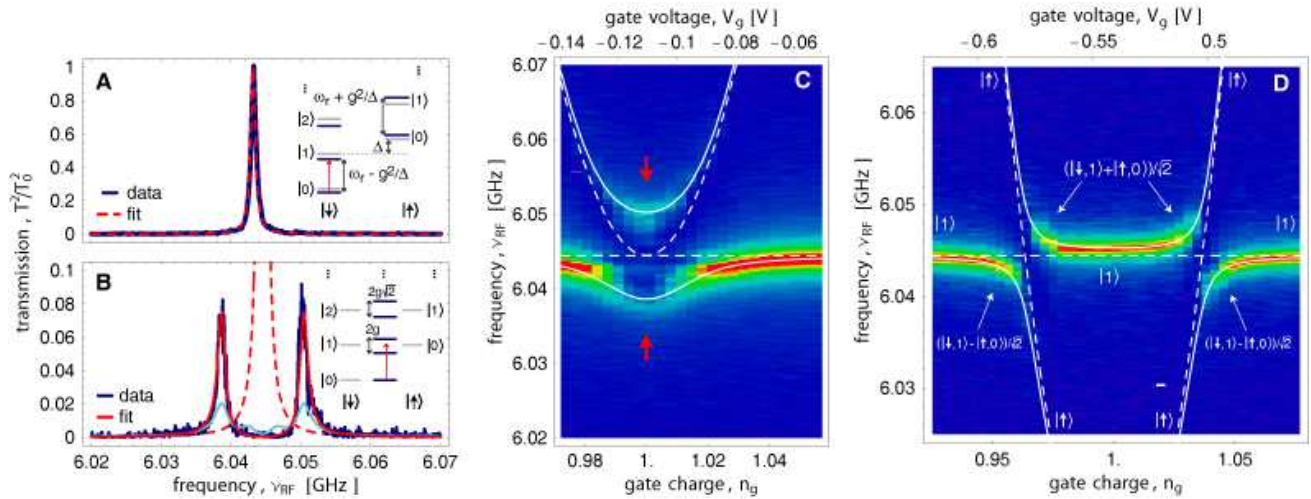}
\end{center}
\caption{Vacuum Rabi mode splitting. \textbf{a} Measured
transmission $T^2$ (blue line) versus microwave probe frequency
$\nu_{\rm{RF}}$ for large detuning ($g^2/\Delta\kappa \ll 1$) and
fit to Lorentzian (dashed red line). The peak transmission
amplitude is normalized to unity. The inset shows the dispersive
dressed states level diagram. \textbf{b} Measured transmission
spectrum for the resonant case $\Delta = 0$ at $n_g = 1$ (blue
line) showing the Vacuum Rabi mode splitting compared to
numerically calculated transmission spectra (red and light blue
lines) for thermal photon numbers of $n =0.06$ and $0.5$,
respectively. The dashed line is the calculated transmission for
$g = 0$ and $\kappa/2\pi = 0.8 \, \rm{MHz}$. The inset shows the
resonant dressed states level diagram. \textbf{c} Resonator
transmission amplitude $T$ plotted versus probe frequency
$\nu_{\rm{RF}}$ and gate charge $n_g$ for $\Delta = 0$ at $n_g=1$.
Dashed lines are uncoupled qubit level separation $\nu_a$ and
resonator resonance frequency $\nu_r$. Solid lines are level
separations found from exact diagonalization of $H_{\rm{JC}}$.
Spectrum shown in b corresponds to line cut along red arrows.
\textbf{d} As in c, but for $E_J/h < \nu_r$. The dominant
character of the corresponding eigenstates is indicated.}
\label{fig:vacuumrabi}
\end{figure*}

\section{The resonant regime}
Making use of the full control over the qubit Hamiltonian, we tune
the flux bias $\Phi_b$ such that the qubit is at $n_g = 1$ and in
resonance with the resonator. Initially, the resonator and the
qubit are cooled into their combined ground state
$\left|0,\downarrow\right\rangle$, see inset in
Fig.~\ref{fig:vacuumrabi}b. Due to the coupling, the first excited
states become a doublet. We probe the energy splitting of this
doublet spectroscopically using a microwave probe beam which
populates the resonator with much less than one photon on average.
The intra resonator photon number is calibrated measuring the
ac-Stark shift of the qubit level separation in the dispersive
case when probing the resonator at its maximum transmission
\cite{schuster04}. The resonator transmission $T^2$ is first
measured for large detuning $\Delta$ with a probe beam populating
the resonator with a maximum of $n \approx 1$ at resonance, see
Fig.~\ref{fig:vacuumrabi}a. From the Lorentzian line the photon
decay rate of the resonator is determined as $\kappa/ 2\pi = 0.8
\, \rm{MHz}$. The probe beam power is subsequently reduced by $5
\, \rm{dB}$ and the transmission spectrum $T^2$ is measured in
resonance ($\Delta = 0$), see Fig.~\ref{fig:vacuumrabi}b. We
clearly observe two well resolved spectral lines separated by the
vacuum Rabi frequency $\nu_{\rm{Rabi}} \approx 11.6 \, \rm{MHz}$.
The individual lines have a width determined by the average of the
photon decay rate $\kappa$ and the qubit decay rate $\gamma$. The
data is in excellent agreement with the transmission spectrum
numerically calculated from the Jaynes-Cummings Hamiltonian
characterizing the qubit decay and dephasing by the single
parameter $\gamma/2\pi = 0.7 \, \rm{MHz}$.

In fact, the transmission spectrum shown in
Fig.~\ref{fig:vacuumrabi}b is highly sensitive to the thermal
photon number in the cavity. Due to the anharmonicity of the
coupled atom-cavity system in the resonant case, an increased
thermal photon number would reduce transmission and give rise to
additional peaks in the spectrum due to transitions between higher
excited doublets\cite{Rau04}. The transmission spectrum calculated
for a thermal photon number of $n = 0.5$, see light blue curve in
Fig.\ref{fig:vacuumrabi}b, is clearly incompatible with our
experimental data. The measured transmission spectrum, however, is
consistent with the expected thermal photon number of $n \lesssim
0.06$ ($T<100\,\rm{mK}$), see red curve in
Fig.\ref{fig:vacuumrabi}b, indicating that the coupled system has
in fact cooled to near its ground state.

Similarly, multiphoton transitions between the ground state and
higher excited doublets can occur at higher probe beam powers.
Such transitions also probe the nonlinearity of the cavity QED
system. In our experiment these additional transitions are not
resolved individually, but they do lead to a broadening of vacuum
Rabi peaks with power. Eventually, at high drive powers the
transmission spectrum becomes single peaked again as transitions
between states high up in the dressed states ladder are driven.

We also observe the anti-crossing between the single photon
resonator state and the first excited qubit state as the qubit is
brought into resonance with the resonator at $n_g = 1$ when tuning
the gate charge and weakly probing the transmission spectrum, see
Fig.~\ref{fig:vacuumrabi}c. For small deviations from $n_g=1$ the
vacuum Rabi peaks evolve from a state with equal weight in the
photon and qubit, as shown in Fig.~\ref{fig:vacuumrabi}b, to
predominantly photon states for $n_g \gg 1$ or $n_g \ll 1$.
Throughout the full range of gate charge the observed peak
positions agree well with calculations considering the qubit with
level separation $\nu_a$, a single photon in the resonator with
frequency $\nu_r$ and a coupling strength of $g/2 \pi$, see solid
lines in Fig.~\ref{fig:vacuumrabi}c.

We have performed the same measurement for a different value of
flux bias $\Phi_b$ such that $E_a/h < \nu_r$ at $n_g = 1$. The
corresponding plot of the transmission amplitude shows two
anti-crossings as the qubit is brought from large positive
detuning $\Delta$ through resonance and large negative detuning
once again through resonance to positive detuning, see
Fig.~\ref{fig:vacuumrabi}d. Again, the peak positions are in
agreement with the theoretical prediction.

\section{Discussion}

We have for the first time observed strong coupling cavity quantum
electrodynamics at the level of a single qubit and a single photon
in an all solid state system. We have demonstrated that the
characteristic parameters of our novel circuit QED architecture
can be chosen at will during fabrication and can be fully
controlled in situ during experiment. This system will open new
possibilities to perform quantum optics experiments in solids. It
also provides a novel architecture for quantum control and quantum
measurement of electrical circuits and will find application in
superconducting quantum computation, for example as a single-shot
QND read-out of qubits and for coupling of qubits over centimeter
distances using the resonator as a quantum bus. This architecture
may readily be used to control, couple and read-out two qubits and
can be scaled up to more complex circuits \cite{Blais04}.
Non-radiative contributions to decoherence in solid state qubits
can be investigated in this system by using the resonator to
decouple the qubit from fluctuations in its electromagnetic
environment. Finally, our circuit QED setup may potentially be
used for generation and detection of single microwave photons in
devices that could be applied in quantum communication.

\begin{acknowledgments}
We thank John Teufel, Ben Turek and Julie Wyatt for their
contributions to the project and are grateful to Peter Day, David
DeMille, Michel Devoret, Sandy Weinreb and Jonas Zmuidzinas for
numerous conversations. This work was supported in part by the
National Security Agency (NSA) and Advanced Research and
Development Activity (ARDA) under Army Research Office (ARO)
contract number DAAD19-02-1-0045, the NSF ITR program under grant
number DMR-0325580, the NSF under grant number DMR-0342157, the
David and Lucile Packard Foundation, the W.~M.~Keck Foundation,
and the Natural Science and Engineering Research Council of Canda
(NSERC).
\end{acknowledgments}

\end{document}